\documentclass[aps,prb,preprint,superscriptaddress,endfloats]{revtex4} 

\usepackage{graphicx}
\usepackage{float}
\usepackage[intlimits,sumlimits,namelimits]{amsmath}
\usepackage{amssymb}
\usepackage{amsfonts,amsthm,amsmath}

\usepackage{bm}

\def\RR{\mathbb R}
\def\TT{\mathbb T}
\def\CC{\rm \hbox{C\kern-.57em\raise.47ex
         \hbox{$\scriptscriptstyle |$}\kern+0.5 em }}

\begin{document}

\title{Some improvements of the ART method for finding transition pathways on potential energy surfaces}

\vspace{0.3in}
\author{E. Canc\`es}
\affiliation{CERMICS, Ecole des
Ponts, Universit\'e Paris-Est, 6 et 8 avenue Blaise Pascal, 77455 Marne-la-Vall\'ee Cedex 2, France}
\affiliation{INRIA Rocquencourt, MICMAC Team-Project, 78153 Le Chesnay
  Cedex, France}
\author{F. Legoll}
\affiliation{Institut Navier, LAMI, Ecole des
Ponts, Universit\'e Paris-Est, 6 et 8 avenue Blaise Pascal, 77455
Marne-la-Vall\'ee Cedex 2, France} 
\affiliation{INRIA Rocquencourt, MICMAC Team-Project, 78153 Le Chesnay
  Cedex, France}
\author{M.-C. Marinica}
\affiliation{CEA, DEN, Service de Recherches de M\'etallurgie Physique, 
91191 Gif-sur-Yvette, France}
\author{K. Minoukadeh}
\affiliation{CERMICS, Ecole des
Ponts, Universit\'e Paris-Est, 6 et 8 avenue Blaise Pascal, 77455 Marne-la-Vall\'ee Cedex 2, France}
\affiliation{INRIA Rocquencourt, MICMAC Team-Project, 78153 Le Chesnay
  Cedex, France}
\author{F. Willaime}
\affiliation{CEA, DEN, Service de Recherches de M\'etallurgie Physique, 
91191 Gif-sur-Yvette, France}

\date{\today}

\begin{abstract}
The Activation-Relaxation Technique $nouveau$ (ART$n$)
is an eigenvector following method for systematic search of saddle points and 
transition pathways on a given potential energy surface. 
We propose a variation of this method aiming at improving the efficiency of 
the local convergence close to the saddle point. We prove the convergence 
and robustness of this new algorithm. The efficiency of the method is tested 
in the case of point defects in body centered cubic iron. 
\end{abstract}
\maketitle

\section{Introduction}

The Activation-Relaxation Technique (ART)
\cite{BM96,BM98,MB98,BM99,MB99,MB00,BM01} is a powerful method for searching
saddle points and transition pathways of a given potential energy surface
(PES). Search methods for saddle points and transition pathways can
actually be classified in 
two main categories. In the first class of methods, one assumes that two
local minima of the PES are known. The main objective of the methods in this
class is to find
the minimum energy path to go from one local minimum to the
other one. The 
Replica Chain method \cite{EK87,AS97}, the Nudged Elastic Band
\cite{MJ94,neb_revue,HJ00}, 
the String method \cite{e_ren_eve,string_ext}, the Transition Path Sampling
\cite{dellago1,dellago2,dellago3,dellago4} and the Discrete Path
Sampling \cite{W02} are some 
methods belonging to this class (note that the Nudged Elastic Band
method has been generalized to the finite temperature setting
\cite{crehuet}, as well as the String method \cite{string_Tpos}). In
the second class of 
methods, one assumes that {\em only one} local minimum of the PES is
known. The aim of methods in this class is to find a saddle point of the PES,
from which the exploration will be pursued toward a different local
minimum, yielding a transition path. Probably the first method in that
class is the EigenVector 
Following method \cite{CM81}. The Dimer method \cite{HJ99}, the
Conformational Flooding method \cite{grubmuller}, the
Hyperdynamics method \cite{hyper1,hyper2,miron}, the Parallel Replica method
\cite{replica}, the Temperature 
Accelerated method \cite{tad1,tad3,tad2}, the Scaled Hypersphere Search
method~\cite{Ohno1,Ohno2,Ohno3}, are other examples. In this
article, we study the Activation-Relaxation Technique,
which belongs to this second class.
We focus here on the zero temperature case, the so-called {\em ART
nouveau} (ART$n$) method~\cite{MB00,MM00,mousseau}, and do not consider the finite temperature case, the
so-called {\em POP-ART} method~\cite{popart}.   
 
\medskip

The ART method is composed of two main steps, the activation step and the
relaxation step. The activation step consists in moving the system from
a local minimum to a saddle point. The relaxation step consists in
relaxing the system, from the computed saddle point, to another local
minimum. Of course, this relaxation step is very fast (and easy to perform) in
comparison with the activation step. 

\medskip

The activation step itself can be divided into two substeps. The first
substep aims at finding some region of the PES with one direction of
negative curvature, which hopefully contains a first order saddle
point, and that we will call the ``attracting region''. The basic idea for
finding a point on the PES with one direction of negative curvature is
to choose a random vector $r$, and next to repeat the two following
operations: (i)~move the system according to $r$, (ii)~relax the system in the
hyperplane orthogonal to $r$, until a point with one direction
of negative curvature has been found (see section 3 for details). The second substep
consists in finding a saddle point in the reached attracting region. From a numerical
viewpoint, 
these two substeps are of very different nature. In this article, we
focus on the second substep, namely the local convergence to a saddle
point, starting from a configuration with one direction of
negative curvature. 
In section~\ref{sec:algo}, we present a simple, prototypical,
ART-like algorithm, which has better local convergence properties than
existing ones. Loosely speaking, this algorithm is optimal in the
principal direction of negative curvature, but suboptimal in the
transverse directions. This is why this algorithm has to be considered 
as a prototype, on the basis of which more complex numerical strategies
can be elaborated. Some numerical results are reported on in
section~\ref{sec:numerics}, where we consider the problem of vacancy
diffusion in crystalline materials. The numerical results obtained on
this problem demonstrate the efficiency of our approach. We gather in
the Appendix a convergence and robustness analysis of this new algorithm. 

\section{A new type of ART-like algorithms}
\label{sec:algo}

From a mathematical viewpoint, a PES for an isolated molecular system
with $N$ atoms is a function $E \, : \,
\RR^{3N}/G_r \longrightarrow \RR$, $N$ being the number of atoms in the
system and $G_r = \RR^3 \times {\rm SO}(3)$ the group of rigid body
movements which act on $\RR^{3N}$ in the following way: for all   
$g=(x_0,R) \in G_r = \RR^3 \times {\rm SO}(3)$, and for all $X =
(x^1,\cdots,x^N) \in \RR^{3N}$, 
$$
g \cdot X = ( R(x_1-x_0), \cdots , R(x_N-x_0) ). 
$$
This viewpoint takes into account the fact that 
the potential energy $E(X)$ of the system is invariant
upon rigid body movements. In the simulation of the condensed phase,
artificial periodic boundary conditions are usually introduced. In this
case, the system is translation invariant, but not rotation invariant,
and a PES then has to be regarded as a function $E \, : \,
\TT^{3N}/\RR^3 \longrightarrow \RR$, where $\TT^{3N}$ is a $3N$
dimensional torus.

For our purpose, namely for the analysis of ART-like methods, there is
no restriction in assuming that the PES under consideration is a
function $f \, : \, \RR^d \longrightarrow \RR$ with isolated critical
points. For $x \in \RR^d$, we denote by $\nabla f(x)$ the gradient of $f$ at the
point $x$ and by $H(x) = \nabla^2f(x)$ the hessian of $f$ at the point
$x$. For $x \in 
\RR^d$, let $\lambda_1(x) \le \lambda_2(x) \le \cdots \le
\lambda_d(x)$ be the eigenvalues of $H(x)$ counted with their
multiplicity, and let $(v_1(x), \cdots, v_d(x))$ be an orthonormal basis
of associated eigenvectors. 

\medskip

Contrarily to second order methods, such as the one proposed
in Ref. \onlinecite{CM81}, the ART 
method does not rely on a complete knowledge of the spectral
decomposition of the Hessian matrix. Instead, it only makes use of
the direction of negative curvature.
We consider here various modifications of the ART method,
differing from the original ART algorithm by the fact that they also
make use of the associated eigenvalue (i.e. of the curvature itself). A
prototype of such algorithm reads   
\begin{equation} \label{eq:algo1}
x_{k+1} = x_k - \frac{(\nabla
  f(x_k),v_1(x_k))}{\min(\lambda_1(x_k),-\lambda_c)} \, v_1(x_k)  
- \mu_t \, \Pi_{v_1(x_k)^\perp} \nabla f(x_k),
\end{equation}
where $\lambda_c > 0$ and $\mu_t > 0$ are fixed numerical parameters,
and $\Pi_{v_1(x_k)^\perp} = I - (v_1(x_k),\cdot) v_1(x_k)$ is the
orthogonal projector on the hyperplane $v_1(x_k)^\perp$.

\medskip

In order to clarify the behavior of the algorithm
(\ref{eq:algo1}) and 
the role of the numerical parameters $\lambda_c > 0$ and $\mu_t > 0$,
let us consider the simple example of a quadratic function $f$:
\begin{equation}
\label{eq:quad}
f(x) = \frac 1 2 \sum_{j=1}^d \lambda_j |x^j|^2,
\end{equation}
with $x=(x^1,\cdots,x^d)$ and $\lambda_1 \le \lambda_2 \le \cdots \le
\lambda_d$. In this simple case, (\ref{eq:algo1}) reads as a system of
$d$ decoupled scalar equations
\begin{eqnarray*} 
x^1_{k+1} & = & \left( 1 - \frac{\lambda_1}{\min(\lambda_1,-\lambda_c)} \right)
x^1_k, \\
x^j_{k+1} & = & \left( 1 - \mu_t \lambda_j \right) x^j_k, \qquad 2 \le j \le d,
\end{eqnarray*}
yielding
\begin{eqnarray*} 
x^1_k & = & \left( 1 - \frac{\lambda_1}{\min(\lambda_1,-\lambda_c)} \right)^k
x^1_0, \\
x^j_k & = & \left( 1 - \mu_t \lambda_j \right)^k x^j_0, \qquad 2 \le j \le d,
\end{eqnarray*}
where $x_0$ is the initial guess of the algorithm.

\medskip

Assume that all the $\lambda_j$ are different from zero. In this case,
$f$ has a unique critical point (the origin), and the algorithm
converges to this critical point for all choices of the initial
guess if and only if 
$$
\lambda_1 < 0 \quad \mbox{and} \quad 0 < \lambda_j < 2\mu_t^{-1} \quad
\mbox{ for all } 2 \le j \le d.
$$
This means that if the algorithm converges, it will be toward a critical
point with Morse index equal to one (a first order saddle
point). Conversely, if $0$ is a saddle point with Morse index equal to
one (i.e. if $\lambda_1 < 0 < \lambda_2 \le \cdots \le \lambda_d$), the
algorithm will converge to zero if and only if $\lambda_d < 2\mu_t^{-1}$. The
numerical parameter $\mu_t$ controls the convergence in the
hyperplane $x^1=0$. If $\mu_t$ is too small,
convergence will be slow, if $\mu_t$ is too large, the algorithm will be
unstable. Note that if $\lambda_1 < - \lambda_c$, convergence in the
$e_1$ direction (the direction of negative curvature) will be obtained
in a single iteration, while linear convergence will be observed if
$-\lambda_c < \lambda_1 < 0$. The role of the parameter $\lambda_c$ is
to prevent the algorithm, when applied to a non-quadratic energy
landscape, from becoming unstable in the region where $|\lambda_1(x_k)|$
is small.

\medskip

Let us now come back to the case of practical interest when $f$ is the
PES of
some molecular system. As mentioned in the introduction, we focus here
on the local convergence properties and
henceforth assume that the iterates have reached the
neighborhood of a first order saddle point. One can then prove (see
the Appendix) that 
the algorithm (\ref{eq:algo1}) converges to the saddle point,
quadratically in the principal direction of negative curvature, and
linearly in the perpendicular directions. Let us note that
quadratic convergence is obtained under the assumption that
the smallest eigenvalue $\lambda_1(x_k)$ of the hessian matrix $H(x_k)$
and the corresponding eigenvector $v_1(x_k)$ are computed exactly.
However, a key ingredient in ART-like algorithms is that
$\lambda_1(x_k)$ and $v_1(x_k)$ are computed approximately, by
iterative methods. Thus, for instance, the eigenelement
$(\lambda_1(x_k),v_1(x_k))$ can be computed by Lanczos or Arnoldi
methods, which are based on repeated evaluations of matrix-vector
products of the form $H(x_k) \, v$. In turn, such matrix-vector products can
be approximately computed using a finite-difference formula, such as
the first-order formula
\begin{equation}
\label{eq:err_H1}
H(x_k) \, v \approx \frac{1}{\epsilon} \left( \nabla f(x_k+\epsilon v) - \nabla
  f(x_k) \right)
\end{equation}
or the second order formula
\begin{equation}
\label{eq:err_H2}
H(x_k) \, v \approx \frac{1}{2\epsilon} \left( \nabla f(x_k+\epsilon v) 
- \nabla f(x_k-\epsilon v) \right).
\end{equation}
In summary, the eigenelement $(\lambda_1(x_k),v_1(x_k))$ is computed
approximately by repeated evaluations of forces $-\nabla f(y)$ for a
collection of configurations $y$ close to the reference
configuration~$x_k$. However, one can prove (see again the Appendix)
that the algorithm (\ref{eq:algo1}) is robust, in the sense that it can
accomodate approximate evaluations of
$(\lambda_1(x_k),v_1(x_k))$. The price to pay is a lower convergence
rate in the principal direction of negative curvature.

\medskip

The prototypical algorithm (\ref{eq:algo1}) is not far from being
optimal in the direction of negative curvature, even in presence of
numerical errors in the evaluation of $(\lambda_1(x_k),v_1(x_k))$. On
the other hand, it is clearly suboptimal in the transverse directions,
where it behaves as a basic fixed step-size gradient. Improvements can
be obtained by resorting to conjugate gradient, quasi-Newton or
trust-region methods in the transverse direction~\cite{NW00}. It is also 
possible, in principle, to take into account the $p$ lowest eigenvalues
of $H(x_k)$ obtained from the Lanczos or Arnoldi partial diagonalization
procedure, to construct a surrogate function that will provide a better
model for $f$ in the neighborhood of $x_k$. Such improvements of the
current ART-like algorithms will be considered in a future work.

\section{Numerical results: Migration of point defects in $\alpha$-iron}
\label{sec:numerics}

In this section we discuss the practical implementation of algorithm
(\ref{eq:algo1}) in the case of basic defects in $\alpha$-iron: small
self interstitial (SIA) and vacancy  (VAC) clusters (1 to 3 defects).
The crystal of $\alpha$-Fe is modeled by the EAM potential developed by
Mendelev {\em et al.}~\cite{mendelev2003,mendelev2004} which has been
the most widely used in recent years to study interstitial loops
~\cite{terentyev2007,terentyev2008}. 
Marinica {\em et al.} have previously used the same potential and the
standard ART$n$ method to test and reveal the energy landscape of small
interstitial clusters (1 to 4 self-interstitials) in $\alpha$-Fe
\cite{marinica}.  It therefore gives us a good basis for comparison.
The crystal consists of 1024$\pm$$n$ atoms ($n$=1,2,3). 
  
\medskip

Starting from a local minimum configuration, the first stage of the
activation step is to push the system out of the basin.  In order to do
this, the system is slightly deformed using
\begin{equation}
\label{eq:basin}
x_{k+1} = x_k + \mu_A \Delta x
\end{equation}
where $\Delta x$ is a fixed normalised deformation of the system, which
is for the moment chosen randomly, and $\mu_A$ is a user-defined fixed
step.  Possibilities for well-chosen initial deformations will be
explored in future work.  In this paper we use the defect centered
deformation \cite{marinica} instead of global deformation.   
This means that the random deformation $\Delta x $ is applied only on
the atoms within a certain radius around the defect.  The reason for
this choice is that the efficiency of the algorithm in the defect
centered deformation is independent of the size of the system and
provides the best rate of successful to unsuccessful activation
processes (this will be elaborated on later on in the section).  At each
iteration the system is relaxed in the hyperplane orthogonal to the
direction $\Delta x$.  If, after this relaxation, the lowest eigenvalue
is still positive, we continue the deformation.  As soon as
$\lambda_1(x_k)$ becomes sufficiently negative ($\lambda_1(x_k) <
\lambda_d$ for some threshold $\lambda_d < 0$), we move onto the next
stage of the activation process.  The threshold is used in order not to
be misguided by numerical errors of the eigenvalue calculation.  The
lowest eigenvalue is computed using the Lanczos algorithm with 15
iterations, a small number compared to the size of the Hessian matrix
(recall that $H(x_k) \in \RR^{3N \times 3N}$, where $N$ is the number of
atoms in the system).   

\medskip

Once the system is out of the basin, we begin to move the system towards
the saddle point, in the hope of following the minimum-energy reaction
path.  The previously used method (slightly modified ART$n$)
\cite{marinica} for this stage is:   
\begin{equation}
\label{eq:algo_old}
x_{k+1} = x_k - \frac{\mu_a}{\sqrt{k}} v_1(x_k)
\end{equation}
where $\mu_a$ is a user-defined constant and $1/\sqrt{k}$ ensures that
the step size gets smaller as we approach the saddle point.  The
direction of the eigenvector $v_1$ is chosen such that it points in the
same direction as the force i.e. $(-\nabla f(x_k),v_1(x_k)) > 0$.  This
is then followed by a relaxation in the hyperplane, which is discussed
in the next paragraph.  Algorithm (\ref{eq:algo_old}) was an improvement
to some previous methods \cite{MB00}. However it has several drawbacks.
The constant parameter $\mu_a$ in the algorithm needs to be defined
according to the PES in study, and even so may be suited for some saddle
point searches but not for others (very tightly positioned saddle points
may force $\mu_a$ to be small for the whole system, which may in turn
impede results when the surface becomes relatively smooth).  With
$v_1(x_k)$ unitary and $\mu_a$ fixed, it is clear that decreasing the
step size according to the number of iterations is not ideal.  In fact
it would be better to use the first and second derivative information of
the energy surface.  Taking as a simple example the function
(\ref{eq:quad}) with $d = 2$ (solution $x_*$ at the origin), we can
position ourselves at a point $x_n$ where the displacement from $x_*$ is
in the direction of negative curvature.  The further along this
direction we are positioned, the more iterations would be needed to
approach $x_*$ since algorithm (\ref{eq:algo_old}) would take smaller
and smaller steps.  In this simple case the proposed algorithm
(\ref{eq:algo1}) would jump to the solution in one step.  

\medskip

The minimization of the forces in the orthogonal hyperplane consists of
following a fixed step steepest descent method but with the force
projected onto $v_1(x_k)^{\bot}$, until the forces in this plane are
zero or a maximum number of steps $M$ is reached.  In the case where we
reach a configuration $x_k$ with $\lambda_1(x_k) > 0$, we restart and
depart from the current local minimum and make a displacement in a
different randomly chosen direction (see Ref. \onlinecite{marinica} for
details on relaxation).  The number of minimization steps taken is
limited due to the potentially costly force evaluations.  The success of
the ART$n$ method however relies strongly on good minimization in the
hyperplane. Possible improvements including trust region and conjugate
gradient methods will be explored in future work.   

\medskip

The main contribution of algorithm (\ref{eq:algo1}) is the step taken in
the direction of the negative curvature.  In the numerical results
reported later on in this section, we have implemented this step in the attracting region.  On the other hand, we continue to use equation (\ref{eq:basin}) for leaving the basin and the same algorithm as described in the previous paragraph for the relaxation in the hyperplane.  

\medskip

The efficiency of these ART-type algorithms depends on two main points: the number of force evaluations required during the activation stage and the ratio of successful to unsuccessful searches.  The failure to find a saddle point can be determined in several ways.  If minimization in the hyperplane is not done sufficiently well, the system risks climbing the energy surface too high.  Once settled at a saddle point, it could be one which is not associated with the local minimum where the activation process began.  It could also be the case that we fall on a saddle point where the energy is lower than the starting point, which is an immediate indication that we have overlooked at least one adjacent saddle point of the local minimum and fallen beyond.  Finally, another sign of failure is when relaxation in the hyperplane yields a positive $\lambda_1(x_k)$, in which case we have reached another local minimum.  It remains a challenge to be certain that a saddle point falls in the first of the three categories mentioned.  For the purposes of this study therefore, we will only reject stationary configurations if the energy is below that of the initial local minimum or if we are in fact at another minimum configuration.    

\medskip

Comparisons between algorithms (\ref{eq:algo_old}) and (\ref{eq:algo1}) are done on interstitial and vacancy defects using the parameters shown in Table \ref{tab_params}.  The results are shown in Table \ref{tab_results}.  Over a total number of 1000 successful events, $\left\langle f \right\rangle$ is the average number of force evaluations per activation process and $\eta$ is the ratio of successful events to unsuccessful events.  It can be observed that the proposed algorithm (\ref{eq:algo1}) improves performance by a large margin both in terms of the average number of force evaluations and the proportion of successful events.  The elimination of the constant factor $\mu_a$ in algorithm (\ref{eq:algo_old}) not only makes the algorithm more efficient but also more versatile.  It may be applied to a wide range of potential energy surfaces without the need for parameter manipulation.  

\begin{table}
\begin{center}
\begin{tabular}{lcccccc}
\hline
\hline
  & \multicolumn{2}{c}{\underline{$n$ SIA}}  & & & \multicolumn{2}{c}{\underline{$n$ VAC}}     \\
                      &   Ref. \onlinecite{marinica}    & This work  & & & Ref. \onlinecite{marinica} & This work \\ 
  \hline
  $\lambda_c$          &   - & 0.5 & & & - & 0.5   \\
  $\lambda_d$         &   -2 & -2  & & & -2 & -2 \\
  $\mu_A$              &   0.6 & 0.6 & & & 0.2 & 0.2   \\
  $\mu_a$                &   0.24 & - & & & 0.08 & -   \\
  $M$                    & 18    & 18  & & & 18 & 18  \\
  \hline
\hline
\end{tabular}
\end{center}
\caption{Parameters used in implementation.  The parameter $\mu_A$ is
  taken from studies by Marinica {\em et al.} \cite{marinica}, with
  $\mu_a/\mu_A = 0.4$. 
\label{tab_params}}
\end{table}

\begin{table}
\begin{center}
\begin{tabular}{llcccccc}
\hline
\hline
  number &					& \multicolumn{2}{c}{\underline{SIA}}  & & & \multicolumn{2}{c}{\underline{VAC}}     \\
 of defects  &					&   ART$n$  \cite{marinica}    & This work  & & & ART$n$ \cite{marinica} & This work \\
  \hline
 1 &$\left\langle f \right\rangle$         &   462 & 298 & & & 780 & 291    \\
  &$\eta$                                  &   4.6 & 4.7 & & & 1.8 & 7.9    \\
 
    \hline
 2 &$\left\langle f \right\rangle$         &   548 & 328 & & & 705 & 323    \\
  &$\eta$                                  &   4.2 & 4.4 & & & 2.6 & 7.1    \\
 
  \hline
 3 &$\left\langle f \right\rangle$         &   691 & 320 & & & 667 & 321    \\
  &$\eta$                                  &   2.6 & 4.4 & & & 2.8 & 7.4    \\
 
    \hline
\hline
\end{tabular}
\end{center}
\caption{Comparison of a previous ART$n$ approach \cite{marinica} and
  the algorithm presented in this article for interstitial and vacancy
  defects.   The new algorithm reduces the average number of force
  evaluations ($\left\langle f \right\rangle$) by about 40\% and 55\%
  for the self-interstitial atoms (SIA) and vacancies (VAC) case
  respectively.  In the case of SIA, the ratio of successful to
  unsuccessful searches ($\eta$) is almost constant. However, in the
  case of vacancies, $\eta$ is increased by over 260\%.
\label{tab_results}}
\end{table}

\section*{Aknowledgements} 

This work was partially supported by the Agence Nationale de la
Recherche (LN3M project, contract No. ANR-05-CIGC-0003). The authors are
grateful to Normand Mousseau for helpful discussions. 

\appendix
\section{Appendix: Local convergence analysis}

In this Appendix, we prove that algorithm (\ref{eq:algo1}) is locally convergent, even when
the eigenelement in the direction of negative curvature is approximately
computed.

Let $x_\ast$ such that $\nabla f(x_\ast) = 0$ and 
$\lambda_1(x_\ast) < 0 < \lambda_2(x_\ast) \le \cdots \le \lambda_d(x_\ast)$. We
introduce the notation $v_1^\ast = v_1(x_\ast)$, $\lambda_1^\ast =
\lambda_1(x_\ast)$, $H_\ast = \nabla^2f(x_\ast)$,
$$
e_k = x_k - x_\ast, \quad
z_k = (x_k-x_\ast)\cdot v_1^\ast, \quad 
y_k = \Pi_{{v_1^\ast}^\perp} (x_k-x_\ast).
$$
Note that 
$$
x_k - x_\ast = z_k v_1^\ast + y_k, \quad \text{hence} \quad
|x_k-x_\ast|^2 = |z_k|^2 + |y_k|^2.
$$
In the analysis below, we often use that $z_k = O(|e_k|)$. 

We consider algorithm (\ref{eq:algo1}), where the eigenelement 
$(\lambda_1(x_k),v_1(x_k))$ is now computed
{\em approximately}. The resulting algorithm, that we analyze
below, reads 
\begin{equation} \label{eq:algo2}
x_{k+1} = x_k - \frac{(\nabla
  f(x_k),\tilde{v}_1(x_k))}{\min(\tilde{\lambda}_1(x_k),-\lambda_c)} 
\tilde{v}_1(x_k)  
- \mu_t \Pi_{\tilde{v}_1(x_k)^\perp} \nabla f(x_k),
\end{equation}
where $\tilde{\lambda}_1(x_k)$ and $\tilde{v}_1(x_k)$ are approximations
of $\lambda_1(x_k)$ and $v_1(x_k)$:
$$
\tilde{v}_1(x_k) = v_1(x_k) + \alpha_k, \quad
\tilde{\lambda}_1(x_k) = \frac{\lambda_1(x_k)}{1 + \beta_k},
$$
where the errors $\alpha_k$ and $\beta_k$ are supposed to be small
(i.e. $|\alpha_k| \ll 1$ and $|\beta_k| \ll 1$). We
assume that $| \tilde{v}_1(x_k) | = 1$.  
Note that we have made no assumption
on the hessian matrix $H(x_k)$. Hence, the errors $\alpha_k$ and
$\beta_k$ take into account both a possible approximation in the computation of
$H(x_k)$ (see (\ref{eq:err_H1}) and (\ref{eq:err_H2})), and
an approximate partial diagonalization of this matrix (by a Lanczos or Anoldi
algorithm). 

We assume that $\lambda_1(x_\ast) < -\lambda_c$ and that $x_k$ is close
enough to $x_\ast$ such that $\lambda_1(x_k) \leq -\lambda_c$ for all $k$
sufficiently large. We also
assume that the error $\beta_k$ is small enough such that
$\tilde{\lambda}_1(x_k) \leq -\lambda_c$ for all $k$ sufficiently large.

It follows from (\ref{eq:algo2}) that
\begin{eqnarray}
\nonumber
z_{k+1} &=& z_k - 
\frac{(\nabla f(x_k),\tilde{v}_1(x_k))}{\tilde{\lambda}_1(x_k)}
(\tilde{v}_1(x_k),v_1^\ast)  
- \mu_t (v_1^\ast, \Pi_{\tilde{v}_1(x_k)^\perp} \nabla f(x_k))
\\
&=&
\label{eq:zk}
z_k - 
\frac{(\nabla f(x_k),v_1(x_k) + \alpha_k )}{\lambda_1(x_k)} \, (1 + \beta_k)
(v_1(x_k) + \alpha_k,v_1^\ast) \nonumber \\ & & 
- \mu_t (\Pi_{\tilde{v}_1(x_k)^\perp}v_1^\ast, \nabla f(x_k))
\end{eqnarray}
and
\begin{eqnarray}
\nonumber
y_{k+1} &=& 
y_k - \frac{(\nabla f(x_k),\tilde{v}_1(x_k))}{\tilde{\lambda}_1(x_k)}
\Pi_{{v_1^\ast}^\perp} \tilde{v}_1(x_k)  - 
\mu_t \Pi_{{v_1^\ast}^\perp} \Pi_{\tilde{v}_1(x_k)^\perp} \nabla f(x_k)
\\
&=& 
\label{eq:yk}
y_k 
- \frac{(\nabla f(x_k),v_1(x_k) + \alpha_k)}{\lambda_1(x_k)} \, (1 + \beta_k)
\Pi_{{v_1^\ast}^\perp} \tilde{v}_1(x_k)  - 
\mu_t \Pi_{{v_1^\ast}^\perp} \Pi_{\tilde{v}_1(x_k)^\perp} \nabla f(x_k).
\end{eqnarray}
Assuming that $f$ is $C^2(\RR^d) \cap L^\infty(\RR^d)$ with bounded
first and second derivatives, it holds
\begin{equation} \label{eq:exp_grad}
\nabla f(x_k) = H_\ast (x_k-x_\ast) + O(|x_k-x_\ast|^2)
= \lambda_1^\ast z_k v_1^\ast + H_\ast y_k + O(|e_k|^2).
\end{equation}
Besides, using perturbation theory, one obtains
\begin{eqnarray}
\nonumber
\lambda_1(x_k) &=& \lambda_1^\ast + (v_1^\ast, (\nabla H(x_\ast) \cdot
e_k) \ v_1^\ast) + O(|e_k|^2),
\\
\label{eq:exp_v1}
v_1(x_k) &=& v_1^\ast - \Pi_{{v_1^\ast}^\perp} \left(
  \left(H_\ast-\lambda_1^\ast\right)|_{{v_1^\ast}^\perp} \right)^{-1}
\Pi_{{v_1^\ast}^\perp} (\nabla H(x_\ast) \cdot e_k) \ v_1^\ast
+ O(|e_k|^2).
\end{eqnarray}
From (\ref{eq:exp_grad}) and (\ref{eq:exp_v1}), we deduce
\begin{eqnarray*}
(\nabla f(x_k),v_1(x_k)) &=& \lambda_1^\ast z_k + O(|e_k|^2),
\\
(v_1(x_k),v_1^\ast) &=& 1 + O(|e_k|^2),
\\
\Pi_{\tilde{v}_1(x_k)^\perp} v_1^\ast &=& O(|e_k|) + O(|\alpha_k|), 
\\
\Pi_{{v_1^\ast}^\perp} \tilde{v}_1(x_k) &=& O(|e_k|) + O(|\alpha_k|),
\\
\Pi_{{v_1^\ast}^\perp} \Pi_{\tilde{v}_1(x_k)^\perp} \nabla f(x_k) &=& 
H_\ast y_k +  O(|e_k|^2) + O(|e_k| \, | \alpha_k |).
\end{eqnarray*}
Inserting these equations in (\ref{eq:zk}) and
(\ref{eq:yk}), we obtain 
\begin{eqnarray}
\nonumber
z_{k+1} &=& z_k -
\frac{(\nabla f(x_k),v_1(x_k))}{\lambda_1(x_k)} 
(v_1(x_k) ,v_1^\ast)  
- \mu_t (\Pi_{\tilde{v}_1(x_k)^\perp}v_1^\ast, \nabla f(x_k))
\\
\nonumber
& & - \, \frac{(\nabla f(x_k),\alpha_k)}{\lambda_1(x_k)} \, (1 + \beta_k)
(v_1(x_k) + \alpha_k,v_1^\ast) 
\\
\nonumber
& &
- \, \frac{(\nabla f(x_k),v_1(x_k))}{\lambda_1(x_k)} \, \beta_k
(v_1(x_k) + \alpha_k,v_1^\ast)  
\\ 
\nonumber
& &
- \, \frac{(\nabla f(x_k),v_1(x_k)  )}{\lambda_1(x_k)} \, 
(\alpha_k,v_1^\ast)  
\\
\label{eq:conv_zk}
&=&
O(|e_k|^2) + O(|e_k| \, |\alpha_k|) + O(|e_k| \, |\beta_k|)
\end{eqnarray}
on the one hand, and, on the other hand,
\begin{eqnarray}
\nonumber
y_{k+1} &=&
y_k 
- \frac{(\nabla f(x_k),v_1(x_k))}{\lambda_1(x_k)} 
\Pi_{{v_1^\ast}^\perp} \tilde{v}_1(x_k)  - 
\mu_t \Pi_{{v_1^\ast}^\perp} \Pi_{\tilde{v}_1(x_k)^\perp} \nabla f(x_k)
\\
\nonumber
& & 
- \, \frac{(\nabla f(x_k), \alpha_k)}{\lambda_1(x_k)} \, 
\Pi_{{v_1^\ast}^\perp} \tilde{v}_1(x_k)  
\\ 
\nonumber
& &
- \, \frac{(\nabla f(x_k),v_1(x_k) + \alpha_k)}{\lambda_1(x_k)} \, \beta_k
\\ 
\label{eq:conv_yk}
&=& 
(I - \mu_t H_\ast) y_k + 
O(|e_k|^2) + O(|e_k| \, |\alpha_k|) + O(|e_k| \, |\beta_k|). 
\end{eqnarray}
As $y_k \in {v_1^\ast}^\perp$ and as $H_\ast$ is positive definite on
${v_1^\ast}^\perp$, we get 
$$
\| (I - \mu_t H_\ast)_{| {v_1^\ast}^\perp} \|_2 = 
\max (1 - \mu_t \lambda_2, \mu_t \lambda_d - 1). 
$$ 
Thus, if $\mu_t < 2/\lambda_d$, we infer from (\ref{eq:conv_zk}) and
(\ref{eq:conv_yk}) that 
$$
|x_{k+1}-x_\ast| \le \gamma |x_{k+1}-x_\ast| + O(|x_k-x_\ast|^2)
+ O(|x_k-x_\ast| \, |\alpha_k|) + O(|x_k-x_\ast| \, |\beta_k|), 
$$
with $\gamma = \| (I - \mu_t H_\ast)_{| {v_1^\ast}^\perp} \|_2 < 1$. 
Under the assumption that the errors $\alpha_k$ and $\beta_k$ are
uniformly bounded by a small constant, this proves that algorithm
(\ref{eq:algo2}) locally converges, 
and that the convergence speed is at least linear. 

In the case when the eigenelement $(\lambda_1(x_k),v_1(x_k))$ is exactly
computed, algorithm (\ref{eq:algo2}) reduces to algorithm
(\ref{eq:algo1}). We hence have proved that algorithm (\ref{eq:algo1})
locally converges, and that this convergence is robust with respect to
errors in the computations of the lowest eigenvalue (and the associated
eigenvector) of $H(x_k)$. 

Estimates for the convergence of algorithm (\ref{eq:algo1}) are readily
obtained from (\ref{eq:conv_zk}) and (\ref{eq:conv_yk}), by setting
$\alpha_k = 0$ and $\beta_k = 0$. We obtain
$$
z_{k+1} = O(|e_k|^2) \quad \text{and} \quad 
y_{k+1} = (1 - \mu_t H_\ast) y_k + O(|e_k|^2).
$$
Note that the convergence for $z_k$ (e.g. in the
principal direction of negative curvature) is {\em quadratic}. If errors
are introduced in the computation of the eigenelement
$(\lambda_1(x_k),v_1(x_k))$, the rate of convergence of $z_k$ becomes
{\em linear}, as can be seen in (\ref{eq:conv_zk}).

\end{document}